\documentstyle[12pt,twoside,fleqn,espcrc1]{article}

\title{Recent results from E877 for Au$+$Au collisions at AGS energy}

\author{Kirill Filimonov for the E877 Collaboration:\\[0.3cm]
  J.~Barrette$^5$, R.~Bellwied$^{9}$, 
  S.~Bennett$^{9}$, R.~Bersch$^7$, P.~Braun-Munzinger$^2$, 
  W.C.~Chang$^7$, W.E.~Cleland$^6$, M.~Clemen$^6$, 
  J.~Cole$^4$, T.~M.~Cormier$^{9}$, 
  Y.~Dai$^5$, G.~David$^1$, J.~Dee$^7$, O.~Dietzsch$^8$, M.~Drigert$^4$,
  S.~Esumi$^3$, K.~Filimonov$^{5,3}$, A.~French$^9$, S.C.~Johnson$^7$, 
  J.R.~Hall$^9$, T.K.~Hemmick$^7$, N.~Herrmann$^3$, B.~Hong$^2$, 
  Y.~Kwon$^7$,
  R.~Lacasse$^5$, Q.~Li$^{9}$, T.W.~Ludlam$^1$,
  S.~K.~Mark$^5$, S.~McCorkle$^1$, 
  D.~Mi\'{s}kowiec$^2$,
  E.~O'Brien$^1$,  
  S.~Panitkin$^7$, V.~Pantuev$^7$, P.~Paul$^7$, T.~Piazza$^7$, M.~Pollack$^7$, 
  C.~Pruneau$^9$, Y.J.~Qi$^5$,
  E.~Reber$^4$, M.~Rosati$^5$, 
  S.~Sedykh$^7$, J.~Sheen$^9$, U.~Sonnadara$^6$, J.~Stachel$^3$,
  N.~Starinski$^5$,
  E.M.~Takagui$^8$,  V.~Topor ~Pop$^5$, M.~Trzaska$^7$, 
  S.~Voloshin$^3$,
  T.B.~Vongpaseuth$^7$,
  G.~Wang$^5$, J.P.~Wessels$^3$, C.L.~Woody$^1$, 
  N.~Xu$^7$,
  Y.~Zhang$^7$, C.~Zou$^7$\\ \bigskip
 $^1$ Brookhaven National Laboratory, Upton, NY 11973\\
 $^2$ Gesellschaft f\"ur Schwerionenforschung, 64291 Darmstadt, Germany\\
 $^3$ Universit\"at Heidelberg, 69120 Heidelberg, Germany\\
 $^4$ Idaho National Engineering Laboratory, Idaho Falls, ID 83402\\
 $^5$ McGill University, Montr\'eal, Canada\\
 $^6$ University of Pittsburgh, Pittsburgh, PA 15260\\
 $^7$ SUNY, Stony Brook, NY 11794\\
 $^8$ University of S\~ao Paulo, Brazil\\
 $^9$ Wayne State University, Detroit, MI 48202\\
}

\begin{document}
\maketitle



The E877 experiment was dedicated to the study of hadron distributions
in
Au+Au collisions at the AGS. Here, we will mainly present results
from the analysis of the data obtained from the last data taking run,
which took place in the fall of 1995 and recorded 46 millions Au+Au
events at 11.5 AGeV/c. The improved experimental apparatus and large
statistics allowed to extend our previous  measurements of particle flow
\cite{Barrette:1994xr,Barrette:1997rr,Barrette:1997pt,Barrette:1998bz} 
to $K^+, K^-,\bar{p}$ and $\Lambda$. We will also report our
latest results on two-particle correlations of
like and unlike particles, as well as the measurements of double
differential multiplicities of $\Lambda$-hyperons. 

\section{UPGRADE OF EXPERIMENTAL SETUP}

The E877 experimental setup is discussed in detail in
\cite{Barrette:1994xr,Barrette:1997rr,Barrette:1997pt,Barrette:1998bz,Barrette:1998vk}. 
For the 1995 run, two 
micro-strip silicon beam vertex
detectors (BVER's) were upgraded from single-sided
silicon wafers with one-dimensional pitch of 50 $\mu$m to
double-sided wafers with a 200 $\mu$m pitch in both the 
$x$ and $y$ directions. 
Using these detectors the coordinates of beam particles at the target were
determined with an accuracy of 300 $\mu$m in position and 60 $\mu$rad in
angle.


We also upgraded the spectrometer with tracking chambers upstream of the
magnet. 
Two identical multi-wire
proportional chambers with highly
segmented chevron shaped cathode pad readout 
were instrumented and placed at 2 m and 2.25 m downstream of
the target, 
just in front of
the spectrometer magnet. They provided a precise measurement (about
$300\mu$m resolution) of the bending plane coordinate of the track 
before deflection in the
magnetic field.
 This additional tracking information allowed to improve
the signal-to-background ratio for identification of rare particles,
such as $K^-$ and $\bar{p}$ and to reconstruct the vertices of $\Lambda$
hyperon decays. 

\section{FLOW OF IDENTIFIED PARTICLES}

Anisotropies in the azimuthal distribution of particles, also called
anisotropic (directed, elliptic, etc.) transverse flow, have proven to be
a very useful tool for extracting information about the hot and dense
stage of a heavy-ion collision. 

In our flow analysis, we reconstructed the reaction plane
from the  measurement of the transverse energy distribution \cite{Barrette:1997pt}.
A Fourier expansion method was used to describe the azimuthal
anisotropy in the particle emission \cite{Voloshin:1996mz}. The amplitudes
of the first and second harmonics of the expansion, $v_1$ and $v_2$,
quantify the strength of directed and elliptic flow, respectively.

\begin{figure}[htb!]
\vskip 40mm
\includegraphics{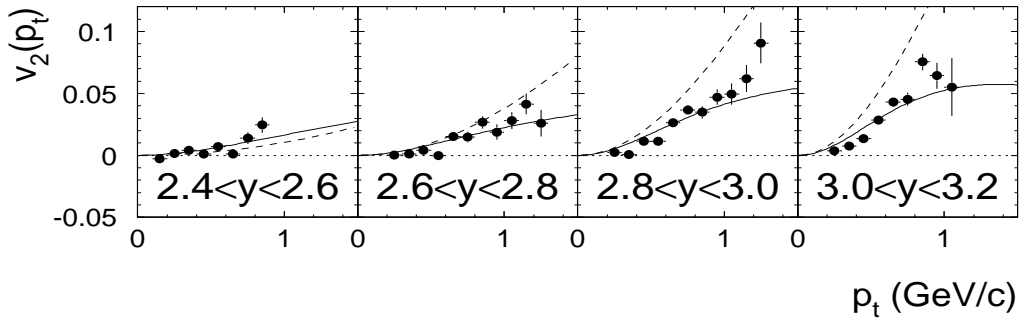}
\vskip -10mm
\caption{\small $v_2(p_t)$ of protons for different rapidity bins and
  centrality $7-12\%~\sigma_{geo}$. The data (solid circles) have been
  corrected for the finite reaction plane resolution.
 Solid and dashed  lines are explained in the text.
}
\label{fig:protons_v2pt}
\vskip -0.6cm
\end{figure}

\begin{figure}[htb]
\vskip 70mm
\begin{minipage}[t]{75mm}
\includegraphics{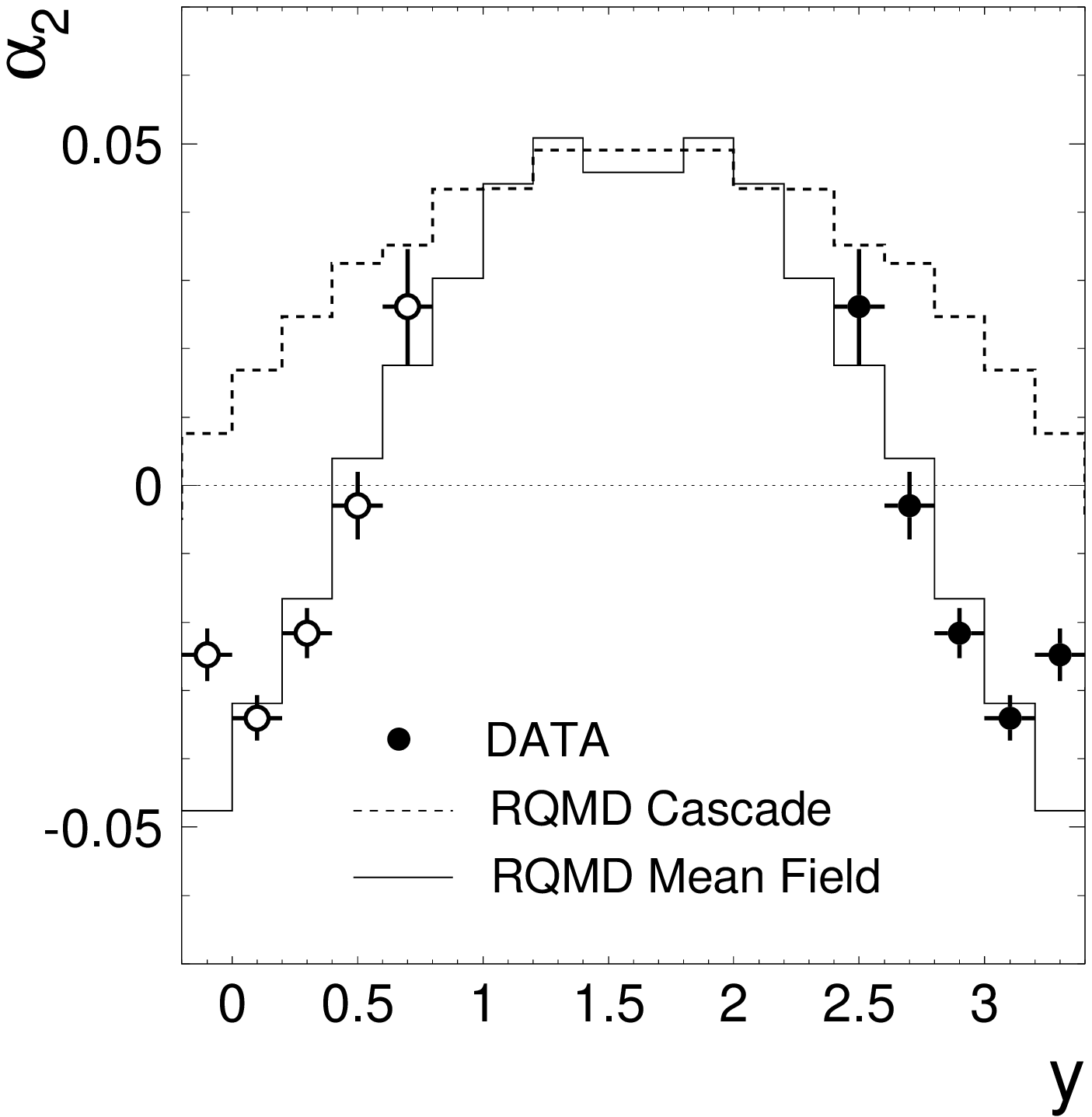}
\vskip -0.5cm
\caption{\small $\alpha_2(y)$ of protons for 
  centrality $26-17\%~\sigma_{geo}$.}
\end{minipage}
\label{fig:protons_v2y}
\hspace{\fill}
\hspace{0.5cm}
\begin{minipage}[t]{75mm}
\includegraphics{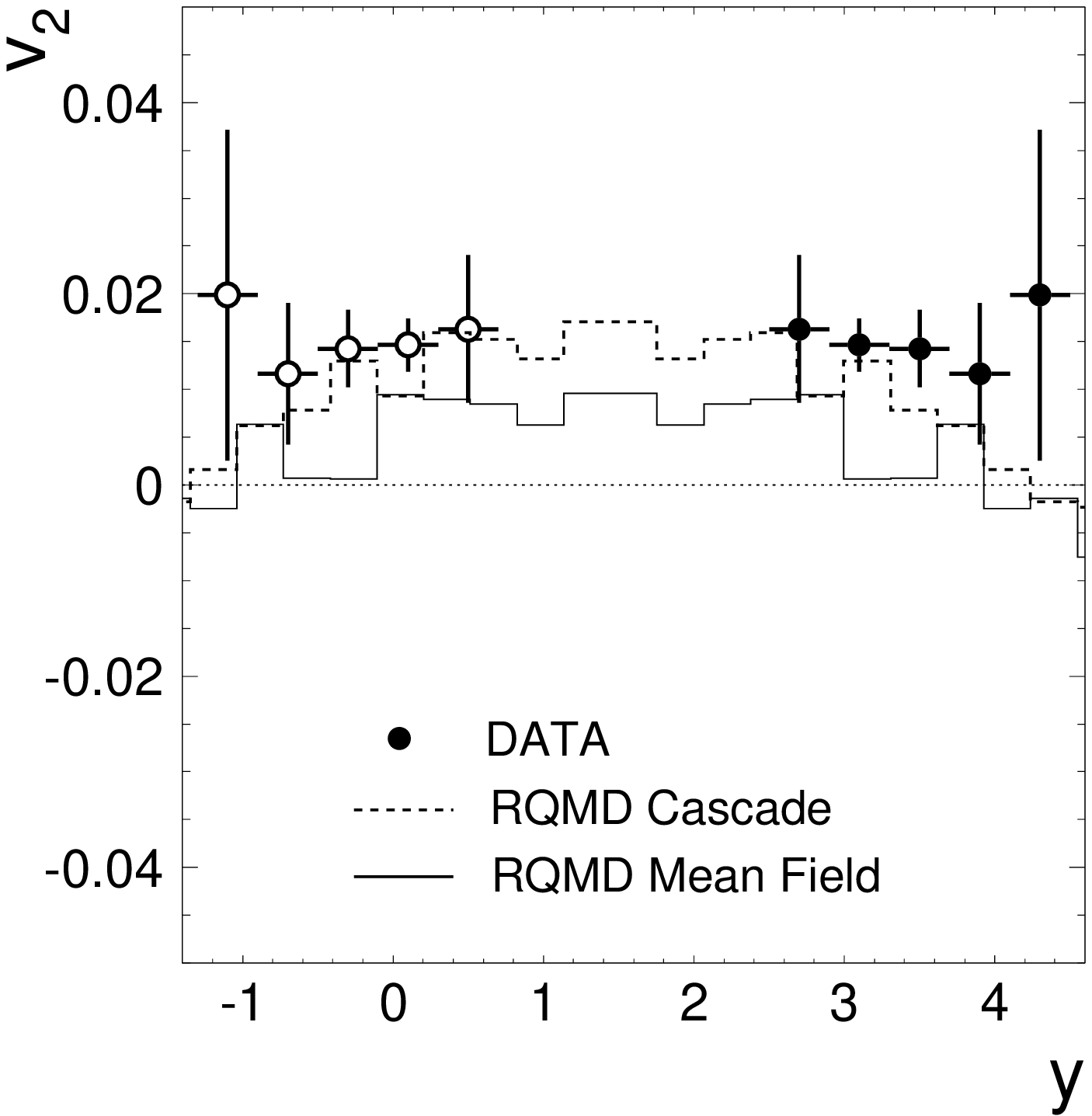}
\vskip -0.5cm
\caption{\small $v_2(y)$ of negative pions for 
  centrality $26-17\%~\sigma_{geo}$.}
\label{fig:pions_v2}
\end{minipage}
\vskip -0.5cm
\end{figure}

Recent theoretical developments have
shown that elliptic flow of nucleons is sensitive to the nuclear
equation-of-state \cite{Danielewicz:1998vz}. 
In Fig.~\ref{fig:protons_v2pt} we show the measured
dependence of $v_2$ on transverse momentum $p_t$ for protons in different rapidity bins 
for events with centrality
$7-12\%~\sigma_{geo}$. Larger positive values of $v_2$
are observed for the more forward rapidities. 
The interpretation of this effect, however, requires considering the influence of a
strong directed flow observed in the region near beam rapidity.  
The directed flow was shown  to be well described 
by a thermal distribution with respect to an origin displaced
along the reaction plane axis \cite{Barrette:1997pt}. The effect of such
a displacement on the
$v_2(p_t)$-dependence is shown in Fig.~\ref{fig:protons_v2pt} by the dashed
lines. To estimate the ``true'' ellipticity of the event, we 
fitted the measured $v_2(p_t)$-dependence with the following parameterization:
\begin{equation}
d^{3}N / d^{3}p \propto
    m_t'\exp(-\frac{m_t'-m}{T_B}),\\
m_t'=\sqrt{(p_x-\langle p_x \rangle)^2+\varepsilon^2\cdot p_y^2+m^2},
\label{eq:mt}
\end{equation}
where $\varepsilon$ is the ellipticity parameter. Values of
$\varepsilon$ larger and less than unity correspond to in-plane and
out-of-plane elliptic flow, respectively. The ellipticity parameter
$\varepsilon$ is related to the second Fourier coefficient $v_2$, 
weighted by $p_t^2$:
\begin{equation}
\alpha_2=v_2^{{\rm weighted~by~} p_t^2}=\frac{\langle p_t^2\cos 2\phi\rangle}{\langle
  p_t^2\rangle}=\frac{\varepsilon^2-1}{\varepsilon^2+1}\sim\varepsilon-1.
\end{equation}
Fits of parameterization (\ref{eq:mt}) to the measured
$v_2(p_t)$-dependence are shown in  Fig.~\ref{fig:protons_v2pt} as solid
lines. The extracted values of $\varepsilon$ for
different centrality and rapidity bins were used to calculate
$\alpha_2$, which is plotted in  Fig.~\ref{fig:protons_v2y} as a
function of rapidity.  
The sign of elliptic flow depends on rapidity, with positive (in-plane)
elliptic flow measured in the midrapidity region, and negative
(out-of-plane) elliptic flow observed at beam rapidities. This
dependence, similar to the transition from out-of-plane to in-plane
elliptic flow measured as a function of beam energy at central rapidities
\cite{Pinkenburg:1999ya},
 could be a consequence of the competition between the out-of-plane flow
 due to shadowing of nucleons by the cold
spectator matter and pressure-induced in-plane elliptic flow caused by
the initial almond-shaped geometry of the collision region. A similar
effect is also seen in
the RQMD calculations (Fig.~\ref{fig:protons_v2y}), with the
mean-field version of RQMD reproducing the data well.

We also studied elliptic flow of pions. Directed flow of pions is very
small \cite{Barrette:1997pt} and does not influence the measurements of
elliptic flow. Pions exhibit a very weak $p_t$-dependence of $v_2$
\cite{jean}, with small positive values observed for semi-central
collisions. 
The rapidity dependence of $v_2$ for
negative pions is presented in Fig.~\ref{fig:pions_v2}. Weak in-plane
elliptic flow is detected at all rapidities covered. In contrast to
protons, the cascade mode of RQMD seems to be in a better agreement with
the pion data.

\begin{figure}[htb]
\vskip 50mm
\includegraphics{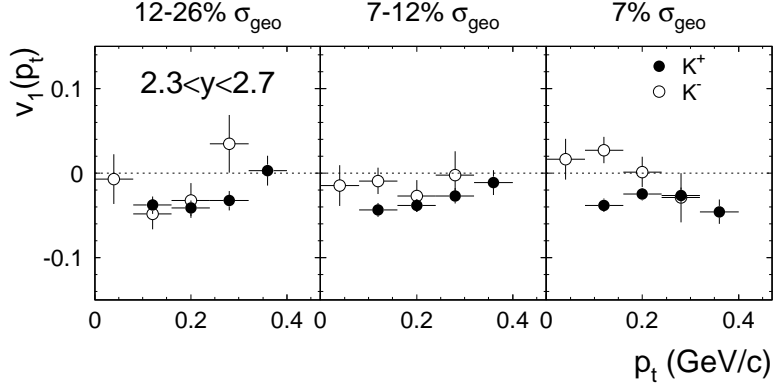}
\hspace*{10.9cm}
{\begin{minipage}[c]{5cm}
\vspace*{-7.5cm}
\caption{\small $v_1(p_t)$ of $K^+$ (solid circles) and $K^-$ (open circles)
  for rapidity $2.3<y<2.7$ and different centralities.}
\label{fig:kaons_v1}
\end{minipage}}
\vskip -1cm
\end{figure}

Azimuthal anisotropies in kaon production are predicted
to be sensitive probes of the kaon potential in a dense nuclear medium 
\cite{Li:1995vd}. 
In Fig.~\ref{fig:kaons_v1} we
present the amplitude of directed flow, $v_1$, as a function of $p_t$
for positive and negative kaons.
The kaon flow signal is weak and negative, i.e. kaons flow in the
direction opposite to that of protons and $\Lambda$-hyperons \cite{jean}.  

\begin{figure}[htb]
\vskip 50mm
\includegraphics{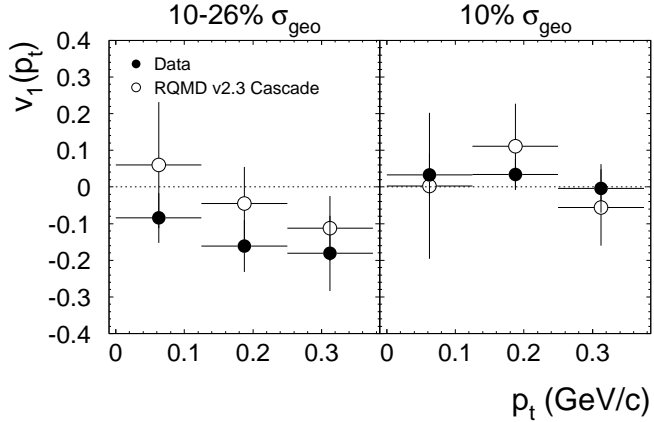}
\hspace*{9.9cm}
{\begin{minipage}[c]{6cm}
\vspace*{-7.5cm}
\caption{\small $v_1(p_t)$ of antiprotons 
  for rapidity $1.8<y<2.2$ and different centralities.}
\label{fig:pbar_v1}
\end{minipage}}
\vskip -0.9cm
\end{figure}

Measurements of directed flow of antiprotons are 
important for understanding the
mechanism of the anti-quark production and the role of
annihilation in the dense environment. Strong anti-flow of
antiprotons has been predicted \cite{Jahns:1994pc} due to absorption of
a large fraction of the produced antiprotons. In Fig.~\ref{fig:pbar_v1}
we compare the experimental measurements of $\bar{p}$ directed flow with
the cascade RQMD calculations.  
A negative signal is observed in semi-central collisions, comparable in
magnitude to that predicted by the model. 

\section{TWO-PARTICLE CORRELATIONS}

Two-particle correlations provide information about the space-time
extent of the system at freeze-out. The correlations of identical mesons
are governed by the Bose-Einstein statistics and by mutual Coulomb
interaction. The two-proton correlations are a product of Fermi statistics
and the attractive
strong and repulsive Coulomb final state interactions. Unlike-pion
and pion-proton correlations are also used as probes of the reaction
zone.

We determined three-dimensional correlation
functions of identical pions to extract the source dimensions
in different directions. The correlation function $C_2$, defined as the ratio
of the two-particle density and the product of the single particle
densities, has been corrected for the repulsive Coulomb force by
integrating the Coulomb wave function over a source of finite size
and fitted according to the following parameterizations:
\begin{equation}
C_2(q_o,q_s,q_l)=1+\lambda e^{-R^2_o
      q^2_o-R^2_s q^2_s-R^2_l q^2_l-2R^2_{ol}q_o q_l} ~~~~{\rm (Bertsch-Pratt)}
\label{eq:bp}
\end{equation}
\begin{equation}
C_2(q_{\perp},q_{\parallel},q_{\emptyset})=1+\lambda e^{-R^2_{\perp}
      q^2_{\perp}-R^2_{\parallel} \gamma^2
      (q_{\parallel}-vq_{\emptyset})^2-\Delta
      \tau^2\gamma^2(q_{\emptyset}-v q_{\parallel})^2} ~~~~{\rm (Yano-Koonin)},
\label{eq:yk}
\end{equation}
\begin{figure}[htb]
  \vskip 98mm
\begin{minipage}[t]{79mm}
    \includegraphics{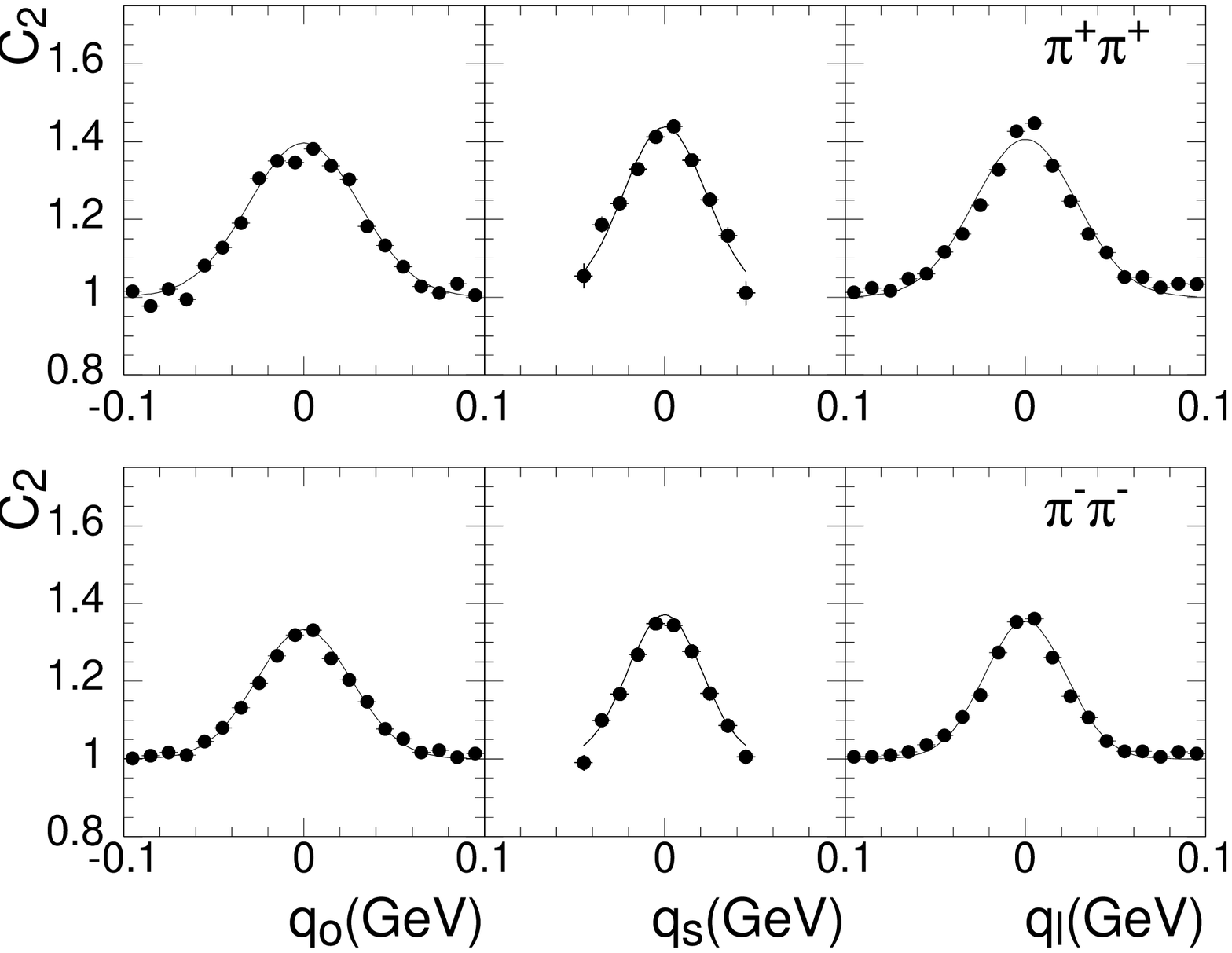}
\end{minipage}
\hspace{\fill}
\begin{minipage}[t]{79mm}
    \includegraphics{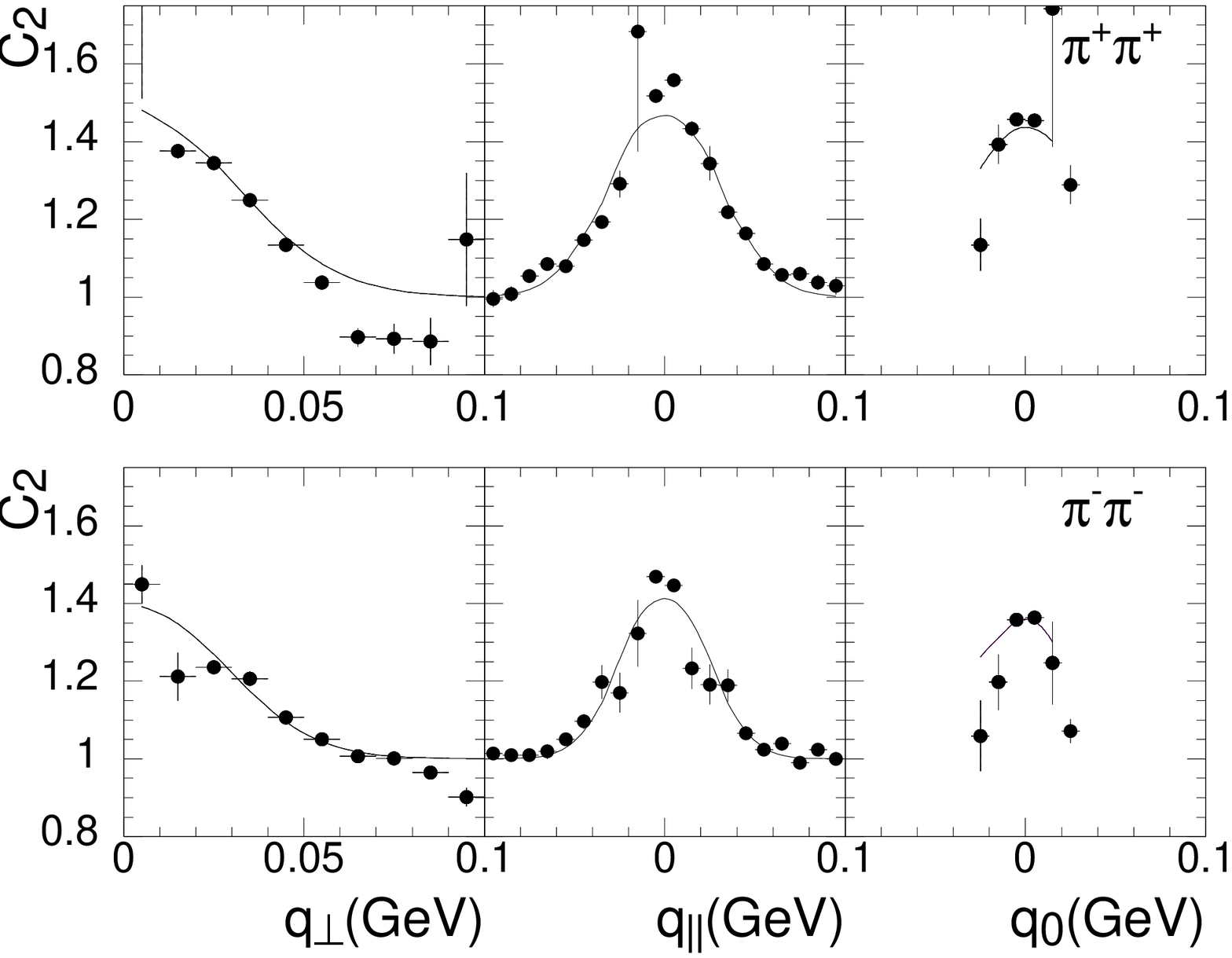}
\end{minipage}
\hspace{\fill}
\vskip -35mm
\vskip -0.5cm
\caption{\small Projections of the Coulomb corrected three-dimensional
  $\pi^+\pi^+$ and  $\pi^-\pi^-$ $C_2(q_o,q_s,q_l)$ (left) and
  $C_2(q_{\perp},q_{\parallel},q_{\emptyset})$ (right) 
for the top $10\%~\sigma_{geo}$ events. For each projection, the other
axes have been integrated from -30 to 30 MeV.}
\label{fig:pion_cor}
\vskip -0.5cm
\end{figure}
where $q_o,q_s,q_l$ and $q_{\perp},q_{\parallel},q_{\emptyset}$ are the
components of the invariant length of the pair's momentum difference.
In Fig.~\ref{fig:pion_cor} we show the projections of the
three-dimensional $\pi^+\pi^+$ and  $\pi^-\pi^-$ $C_2$'s
for the $10\%$ most central events.
The results of the fits
are summarized in Table~\ref{tab:pion}.
\begin{table}[hbt]
\newlength{\digitwidth} \settowidth{\digitwidth}{\rm 0}
\catcode`?=\active \def?{\kern\digitwidth}
\caption{\small Fit parameters to $C_2(q_o,q_s,q_l)$ and
  $C_2(q_{\perp},q_{\parallel},q_{\emptyset})$ of $\pi^+\pi^+$ and
  $\pi^-\pi^-$ pairs for $10\%~\sigma_{geo}$.}
\label{tab:pion}
\begin{tabular*}{\textwidth}{@{}@{\extracolsep{\fill}}l@{\extracolsep{\fill}}c@{\extracolsep{\fill}}cl@{\extracolsep{\fill}}c@{\extracolsep{\fill}}c}
\hline
$C_2(q_o,q_s,q_l)$& $\pi^+\pi^+$ & $\pi^-\pi^-$ &
$C_2(q_{\perp},q_{\parallel},q_{\emptyset})$& $\pi^+\pi^+$ & $\pi^-\pi^-$\\ 
\hline 
$R_o$(fm) & $4.63\pm 0.08$ & $5.31\pm 0.05$ & $R_\perp$(fm) & $4.79\pm 0.07$ & $5.45\pm 0.05$\\
$R_s$(fm) & $6.04\pm 0.15$ & $6.76\pm 0.10$ & $R_\parallel$(fm) & $4.22\pm 0.11$ & $5.85\pm 0.07$\\
$R_l$(fm) & $5.00\pm 0.09$ & $6.18\pm 0.06$ & $\Delta\tau$(fm) &
$0.00\pm 0.39$ & $0.00\pm 0.37$\\
$R^2_{ol}$(fm$^2)$ & $7.19\pm 0.60$ & $6.66\pm 0.45$& $Y_{source}$& $2.37\pm 0.03$ & $2.64\pm 0.02$ \\
$\lambda$ & $0.61\pm 0.01$ & $0.58\pm 0.01$ & $\lambda$ & $0.60\pm 0.01$
& $0.57\pm 0.01$ \\
\hline
\end{tabular*}
\end{table}
The averaged pair rapidities (transverse momenta) were 3.18 (0.14
GeV/c) and 3.34 (0.11 GeV/c) for $\pi^+\pi^+$ and  $\pi^-\pi^-$,
respectively.

The extracted radii of the pion source demonstrate that the system
significantly expanded before freeze-out from its initial size
($R_{Au}^{rms,1-dim}=3.1$ fm).
The $\pi^-\pi^-$ radius parameters are about
$10\%$ larger than those of $\pi^+\pi^+$. This effect is
consistent with the influence of a central Coulomb potential of a third
body with $Z_{eff}\sim 150$ \cite{Baym:1996wk}.

The cross-term parameter $R_{ol}$ in
parameterization (\ref{eq:bp}) is different from zero indicating that
the pion source is not completely boost invariant. The other two cross-terms $R_{sl}$ and $R_{os}$  were found to be consistent with zero
which is expected for an azimuthally symmetric
collision region. The values of $Y_{source}$ confirm that the fastest
pions source is about one unit forward of mid-rapidity.

A possible anisotropy in coordinate space was also studied by measuring
 one-di\-men\-si\-onal $\pi^+\pi^+$ and  $\pi^-\pi^-$
correlation functions relative to the reaction plane. 
For events with centrality 
$9-16\%~\sigma_{geo}$ we divided the pion data set into four subsets
depending on the direction $\phi_{K_t}$ of the pair's transverse momentum  with
respect to the reaction plane angle $\psi_R$: ``same'' with
$\cos(\phi_{K_t}-\psi_R)>1/\sqrt{2}$; ``opposite'' with
$\cos(\phi_{K_t}-\psi_R)<-1/\sqrt{2}$; ``in-plane'' with
$|\cos(\phi_{K_t}-\psi_R)|>1/\sqrt{2}$; ``out-of-plane'' with
$|\cos(\phi_{K_t}-\psi_R)|<1/\sqrt{2}$. The ratios of correlation functions
for pion pairs in ``same''/''opposite''
and ``in-plane''/''out-of-plane'' windows were found to be consistent
with unity, indicating no or very weak anisotropy in the coordinate
space. The upper limits for the azimuthal asymmetry are $8\%$ and $6\%$
 for the $\pi^+$ and $\pi^-$ sources, respectively \cite{thongbay}.

For the analysis of the $K^+K^+$ correlation
function, due to limited statistics we used a two-dimensional parameterization: $C_2(q_t,q_l)=1+\lambda e^{-R^2_t
      q^2_t-R^2_l q^2_l}$. The average rapidity (transverse momentum) of
    the $K^+K^+$ pair was 2.34 (0.16 GeV/c). 
 The extracted parameters of the kaon source are
    $R_t=3.04\pm 0.40$ fm, $R_l=2.73\pm 0.47$ fm, and $\lambda=0.86\pm 0.13$.
The smaller source size inferred from the $K^+K^+$ correlations may be
interpreted as a consequence of the large $m_t$ and 
collective expansion of the system.

We also performed a study of two-proton correlations in Si+Pb and 
Au+Au collisions which
suggested a larger space-time extent for the heavier system. For Au+Au
reactions, the centrality dependence of the 
two-proton correlation function implies that more central collisions
lead to larger source sizes \cite{Barrette:1999qn}.

Measurements of the pion-proton correlation function exhibit an 
asymmetric peak, which could be a sign of spatial or temporal separation
between the sources of protons and pions. Calculations assuming that the
proton source is located 10 fm more forward than the sources of pions,
while the $\pi^+$ and $\pi^-$ sources are separated 
by 15 fm in the direction of the
reaction plane have yielded asymmetries in the calculated 
$\pi^-p$ and $\pi^+p$ correlation
functions which are comparable to the ones 
observed in the experimental data \cite{Miskowiec:1998ms}. The magnitude
of the assumed displacement seems large, though, compared to the size of the
system and the expected duration of the emission. 

\section{$\Lambda$ SPECTRA AND YIELDS}
\begin{figure}[htb!]
  \vskip 85mm
\begin{minipage}[t]{75mm}
    \includegraphics{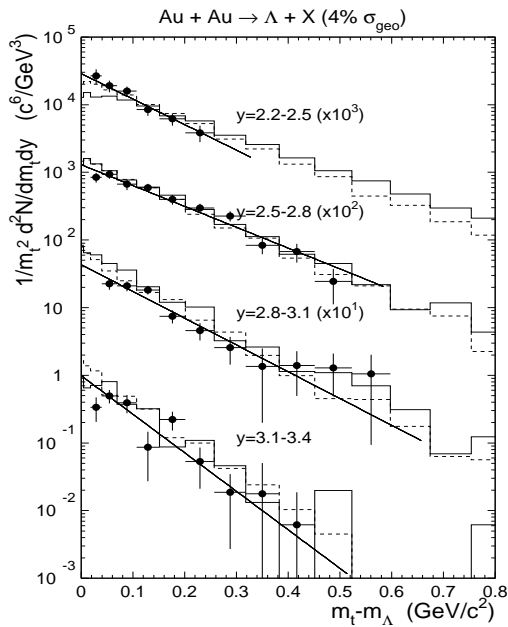}
\end{minipage}
\hspace*{8.4cm}
{\begin{minipage}[c]{7.5cm}
\vspace*{-11.5cm}
\caption{\small $\Lambda$ transverse mass spectra for centrality
  $4\%~\sigma_{geo}$. The data (solid circles) fitted by an exponential
  (solid lines) and
  calculations of the RQMD v2.3 run in cascade (dashed histograms) and
  mean-field (solid histograms) modes are presented in rapidity bins of 0.3
  unit widths successively multiplied by increasing powers of 10.}
\label{fig:lambda_spectra}
\end{minipage}}
\vskip -1cm
\end{figure}

\begin{figure}[htb!]
\vskip 65mm
\includegraphics{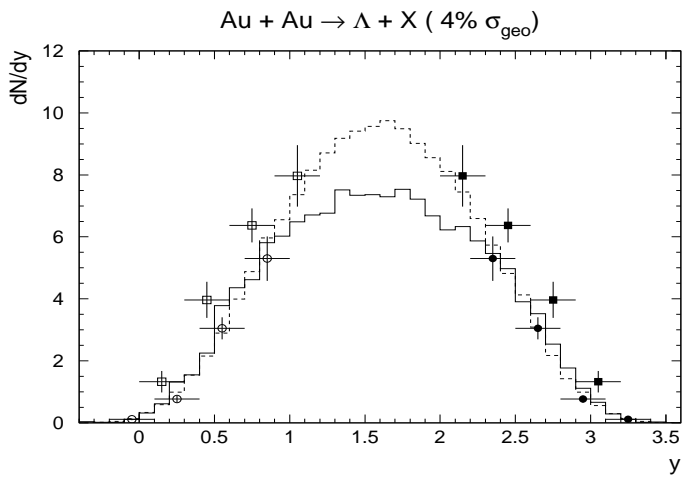}
\hspace*{9.9cm}
{\begin{minipage}[c]{6cm}
\vspace*{-7.5cm}
\caption{\small $\Lambda$ rapidity distribution for centrality $4\%~\sigma_{geo}$.
      The E877 data (circles) and E891 data (squares) are compared with
  the calculations of the RQMD v2.3 run in cascade (dashed histogram) and
  mean-field (solid histogram) modes. The data (solid symbols) are
  reflected about $y_{cm}=1.6$ (open symbols).}
\label{fig:lambda_dndy}
\end{minipage}}
\vskip -0.5cm
\end{figure}

We identified $\Lambda$-hyperons by invariant mass reconstruction of
$(p,\pi^-)$ pairs (see \cite{jean,yujin}). 
The $\Lambda$ spectra in the top $4\%$~$\sigma_{geo}$ 
Au+Au collisions are shown in Fig.~\ref{fig:lambda_spectra}
as a function of transverse mass and rapidity. The spectra are well
described by an exponential with inverse slopes decreasing with increasing
rapidity. The data are compared with the predictions of the RQMD v2.3
model, run in cascade and mean-field modes. In the rapidity
range covered, either version of the model describes the data quite well. The rapidity
distribution for $\Lambda$ obtained by integrating the transverse mass
spectra is shown in Fig.~\ref{fig:lambda_dndy}. Our experimental data
are compared with the measurements of the E891 collaboration
\cite{Ahmad:1991nv} and with the RQMD predictions. Both experimental
results agree within systematic uncertainties. 



\end{document}